\documentclass[iop]{emulateapj}
\usepackage{amssymb,multirow,color,amsmath}

\usepackage{graphicx}
\usepackage{dcolumn}
\usepackage{bm}
\allowdisplaybreaks

\begin{document}

\preprint{}

\title{Spectral breaks of Alfv\'enic turbulence in a collisionless plasma}
 
\author{Stanislav Boldyrev$^1$, Christopher H. K. Chen$^{2}$, Qian Xia$^1$, Vladimir Zhdankin$^1$}
\affiliation{{}$^1$Department of Physics, University of Wisconsin--Madison, Madison, WI 53706, USA \\
{}$^2$Department of Physics, Imperial College London, London SW7 2AZ, UK}
\date{\today}

\begin{abstract}
Recent observations reveal that magnetic turbulence in the nearly colisionless solar wind plasma extends to scales smaller than the plasma microscales, such as ion gyroradius and ion inertial length. Measured breaks in  the spectra of magnetic and density fluctuations at high frequencies are thought to be related to the transition from large-scale hydromagnetic to small-scale kinetic turbulence. The scales of such transitions and the responsible physical mechanisms are not well understood however.  In the present work we emphasize  the  crucial role of the plasma parameters in the transition to kinetic turbulence, such as the ion and electron plasma beta, the electron to ion temperature ratio, the degree of obliquity of turbulent fluctuations. We then propose an explanation for the spectral breaks reported in recent observations.   
\end{abstract}
\maketitle

{\section {Introduction}}
In situ measurements of the magnetic, electric, and density fluctuations in the solar wind provide valuable information on nonlinear dynamics of a nearly collisionless astrophysical plasma. At large hydrodynamic scales (corresponding to low frequencies in the measurements), such fluctuations are thought to be consistent with magnetohydrodynamic turbulence viewed as interacting oblique Alfv\'en modes propagating along the background magnetic field \cite[][]{iroshnikov63,kraichnan65,goldreich_s95,galtier_nnp00,boldyrev06}. At higher frequencies the spectrum of such turbulence exhibits a break, which corresponds to the spatial scale broadly consistent with the plasma microscales such as the ion gyroradius or the ion inertial length. 

It has been proposed that the spectral break in the solar wind and other astrophysical plasmas can mark a transition from the non-dispersive Alfv\'en modes to the dispersive kinetic-Alfv\'en modes  \cite[][]{bale_etal2005,leamon_etal1998,leamon_etal1999,hollweg1999,howes_etal2006,chandran_etal2009,shaikh_zank2009,chandran_etal10,chen_whs2010,howes_q2010,petrosyan_etal2010,howes_td2011,chandran_dqb2011,tenbarge_h2012,boldyrev_p12,sahraoui_etal2012,mithaiwala2012,boldyrev_perez2013,podesta2013,chen_b2013,haverkorn_s2013}. A possibility of transition to whistler turbulence has also been considered   \cite[][]{beinroth1981,coroniti1982,goldstein_rf1994,stawicki_gl2001,galtier_b03,galtier_b05,gary_sl2008,saito_gn2008,gary_s2009,shaikh2009,shaikh2010,gary_sn2010}, however, recent studies   \cite[e.g.,][]{podesta2013,chen_b2013} suggest that whistler turbulence, if present at subpropton scales, contributes only a small fraction of fluctuations energy. A conclusion is then drawn that the spectral break occurs at the proton gyroscale. 

A recent work  by \citet[][]{chen_etal2014} tested this prediction by analyzing the solar wind intervals having very large and  very small plasma beta, the ratio of the kinetic energy of the plasma particles to the magnetic energy. In particular, the intervals were selected with ion and electron plasma beta satisfying $\beta_i\sim \beta_e \gg 1$ and $1\gg \beta_e \gg \beta_i$.  
In the first case, the theory of oblique Alfv\'enic turbulence predicts the break at the ion gyroscale, in the second one at the ion acoustic scale. The observations of \cite[][]{chen_etal2014} agree with the theory in the first case, and disagree in the second, where the break is observed at the ion inertial length instead.  This puzzling result may question the applicability of the theory of turbulent cascade to the solar wind plasma. 

We address this contradiction by inspecting the theory of Alfv\'en turbulence in the limiting cases of large and small plasma beta. 
We consider various mechanisms that may be responsible for the spectral break, including the possibility that the major assumption of the standard theory, the obliquity of propagation, can break down in the case $1\gg \beta_e\gg \beta_i$. The reason for the latter possibility is upscatter of the Alfv\'enic fluctuations due to their interactions with the ion-acoustic modes and the fast modes that are weakly damped in a non-isothermal low-beta plasma.  The Afv\'enic turbulence then develops a non-oblique component $k_\|\gtrsim k_\perp$, which is dissipated due to the ion cyclotron resonance at the ion inertial scale thus explaining the observed spectral break in this case.\\


\section{Kinetic Derivation}
In what follows we will use the notation: $\omega_{p\alpha}=\sqrt{4\pi n_{0\alpha}q^2_{\alpha}/m_\alpha}$ is the plasma frequency, and $v_{T\alpha}=\sqrt{T_\alpha/m_\alpha}$ is the thermal velocity associated with the particles of kind $\alpha$. For a plasma consisting of electrons and ions, the so-called ion-acoustic velocity can be defined, $v_{s}=\sqrt{T_e/m_i}$. It is also convenient to introduce the Alfv\'en speed $v_A=B_0/\sqrt{4 \pi n_0 m_i}$, and the plasma beta, which is the ratio of the thermal energy of the particles to the magnetic energy of the plasma, and which can be different for the ions and the electrons, $\beta_i=2v_{Ti}^2/v_A^2$, $\beta_e=2v_{s}^2/v_{A}^2$. The particle gyroradii are denoted by $\rho_{\alpha}=v_{T\alpha}/\Omega_\alpha$, where $\Omega_\alpha$ are corresponding gyrofrequencies; the ion inertial length is denoted by $d_i=v_A/\Omega_i$.

In order to find the dispersion relations for the waves that can propagate in a plasma, one needs to solve the equation $\det(D_{lm})=0$, where
$D_{lm}\equiv k^2\delta_{lm} -k_lk_m -\frac{\omega^2}{c^2}\epsilon_{lm}(\omega, {\bf k}),$  
and the general expression for dielectric tensor $\epsilon_{lm}$ can be found in \cite[e.g.,][]{stix1992,aleksandrov_br1984,boldyrev_etal2013}. 
Analytic solutions of this equation can be found for the important limiting cases that we consider below in more detail.

\subsection{Spectral breaks at $\beta_i \gg 1$}
In this case the dispersion corrections associated with the ion gyroradius, $\sim k \rho_i$ dominate the corrections associated with the ion inertial length, $\sim k d_i$, and the latter can therefore be neglected. In terms of frequency of Alfv\'en waves this means that $\omega \ll \Omega_i$ and $\omega\ll k_zv_{Ti}$.  Keeping in mind that we will be interested in oblique propagation, we will also assume $k_z\rho_i\ll1$, however, we do not make assumptions about smallness of $k_\perp \rho_i$.   

The condition $\omega\ll \Omega_i$ allows one to simplify the dielectric tensor.  The approximate expressions for the components of $\epsilon_{ij}$ including both electron and ion contributions, to the first order in small $k_z^2 \rho^2_i$ have the form:
\begin{eqnarray}
\label{exx1} \epsilon_{xx}&\approx & \frac{ \omega_{pi}^2}{k_\perp^2 v_{Ti}^2}
\left[1-A_0(z_\perp)+6z_\| \sum\limits_{n=1}^{\infty} \frac{A_n(z_\perp)}{n^2}  \right],\\
\label{eyy1}\epsilon_{yy}& \approx & \epsilon_{xx}-\frac{4\omega_{pi}^2z_\perp}{\Omega_i^2}\sum\limits_{n=1}^{\infty}\frac{A_n^\prime(z_\perp)}{n^2} -\nonumber \\
&-& i\sqrt{2\pi} \frac{\omega_{pi}^2k_\perp^2 v_{Ti}}{\Omega_i^2\omega |k_z|}A^\prime_0(z_\perp), \\
\label{exy1}\epsilon_{xy}&= &-\epsilon_{yx}\approx -i\frac{\omega_{pi}^2}{\omega \Omega_i}\left[1+A_0^\prime(z_\perp) \right],\\
\label{exz1}\epsilon_{xz}&= &\epsilon_{zx}\approx -\frac{2\omega_{pi}^2 k_z}{\Omega^2_i k_\perp}\sum\limits_{n=1}^\infty \frac{A_n(z_\perp)}{n^2},\\
\label{eyz1}\epsilon_{yz}&= &-\epsilon_{zy}\approx -i\frac{\omega_{pi}^2 k_\perp}{\omega\Omega_ik_z}\left[1+A_0^\prime(z_\perp)\right]+\nonumber \\ 
&+& \sqrt{\frac{\pi}{2}}\frac{\omega_{pi}^2 k_\perp}{\Omega_ik_z|k_z|v_{Ti}}A_0^\prime(z_\perp), \\
\label{ezz1}\epsilon_{zz}&\approx & \frac{\omega_{pe}^2}{k_z^2 v_{Te}^2} +   \frac{\omega_{pi}^2}{k_z^2v_{Ti}^2}A_0(z_\perp)+\frac{2\omega_{pi}^2}{\Omega_i^2}\sum\limits_{n=1}^\infty \frac{A_n(z_\perp)}{n^2} + \nonumber \\
&+&i\sqrt{\frac{\pi}{2}}\frac{\omega}{|k_z|v_{Ti}} \frac{\omega_{pi}^2}{k_z^2v_{Ti}^2}A_0(z_\perp).
\end{eqnarray}
In these expressions the coordinate frame is chosen such that the z-direction is along the uniform magnetic field, and the wavevector has the coordinates $(k_\perp, 0, k_z)$. We use the notation $z_\perp=k_\perp^2\rho_i^2$, $z_\|=k_z^2\rho_i^2$,  $A_n(z_\perp)=I_n(z_\perp)\exp(-z_\perp)$, where $I_n$ is a modified Bessel function of order~$n$, and $A^\prime_n(z_\perp)$ denotes the derivative with respect to~$z_\perp$. The last terms in expressions~(\ref{eyy1}), (\ref{eyz1}), and (\ref{ezz1}) describe the collisionless dissipation by ions. As we will see momentarily, the imaginary term in $\epsilon_{yy}$ becomes significant close to the ion gyroscale.  
  
The Bessel functions in the dispersion equation $\det(D_{lm})=0$ can be simplified for large and small arguments. We consider these cases separately. In the case $z_\perp \ll 1$, one can check that for $\omega\approx k_zv_A$ and $\beta_i\gg 1$, the dissipation term provides the dominant contribution in both $\epsilon_{yy}$ and $D_{yy}$, while it can be neglected in~(\ref{ezz1}). In the same limit, the $\epsilon_{yz, zy}$ and $\epsilon_{xz, zx}$  components of the dielectric tensor can be neglected to the leading order in~$1/\beta_i$.

The dispersion equation then takes the form: 
\begin{eqnarray}
\omega^2=\frac{k_\perp^2c^2}{\epsilon_{zz}}+\frac{k_z^2 c^2}{\epsilon_{xx}}+\frac{\omega^2|\epsilon_{xy}|^2}{\epsilon_{xx}\epsilon_{yy}},   
\label{disp_high_beta}
\end{eqnarray}
where the last term describes the dissipation correction, and one can substitute $\omega=k_zv_A$ in this term. As a result we obtain:
\begin{eqnarray}
\omega^2&=&k_z^2v_A^2\left[1+\left(\frac{3}{4}+\frac{T_e}{T_i+T_e} \right)k^2_\perp \rho_i^2 -3k_z^2\rho_i^2 -\right. \nonumber \\
&-& \left. i\frac{9}{8}\sqrt{\frac{\beta_i}{\pi}}k_\perp^2\rho_i^2\right],
\label{large_beta}
\end{eqnarray} 
which is the dispersion relation for Alfv\'en waves in a collisionless high-beta plasma.  This relation holds for the shear-Alfv\'en modes, where the inertial corrections ($\sim kd_i$) are negligibly small. The pseudo-Alfv\'en modes are strongly damped in the considered limit, and they are not discussed here. We note that the dissipation rate, given by the expression
$\gamma/\omega =-(9/16)\sqrt{\beta_i/\pi}\,k_\perp^2\rho_i^2$, 
becomes significant at the ion gyroscale. 

\begin{figure}[tbp]
\includegraphics[width=1.\columnwidth]{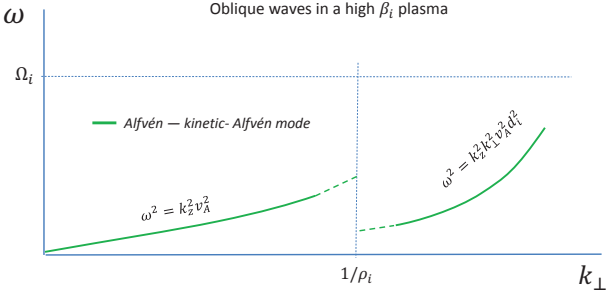}
\caption{\label{fig:curves_high_beta} Sketch of dispersion relations for oblique waves in a collisionless high-$\beta_i$ plasma. The dashed lines indicate the strongly damped modes.}
\end{figure}

For $z_\perp \gg 1$, the non-dissipative components of the tensor $D_{lm}$ agree to leading order with the components of $D_{lm}$ for the subproton kinetic-Alfv\'en waves obtained in  \cite[][Eqs.~(32-37)]{boldyrev_etal2013}. The dominant ion dissipation term comes in this limit through the last term of~(\ref{eyy1}). We thus get the dispersion relation for the high-beta kinetic-Alfv\'en waves:
\begin{eqnarray}
\omega^2=\frac{k_z^2k_\perp^2 v_A^4}{\Omega_i^2}\left[1-\frac{i}{2k_\perp^2\rho_i^2} \right].
\label{large_beta2}
\end{eqnarray}
The dissipation rate estimated from this expression,
$\gamma/\omega=-1/(4k_\perp^2 \rho_i^2)$, 
also becomes significant at the ion gyroscale. The resulting dispersion curve is schematically shown in~Fig.~(\ref{fig:curves_high_beta}). The dispersion and dissipation terms in (\ref{large_beta}, \ref{large_beta2}) suggest that the spectral break occurs close to the ion gyroscale $\rho_i$ independently of the obliquity of propagation.\footnote{The significance of the collisionless ion dissipation in the vicinity of the ion gyroradius has been observed in numerical solution of the Vlasov-Maxwell equations   \cite[e.g.,][]{howes_etal2006,schekochihin_etal2009}. Our formulae (\ref{large_beta}, \ref{large_beta2}) provide analytic explanation for those observations.} This is consistent with the observational measurements \cite[][]{chen_etal2014}.

\subsection{Spectral breaks at $\beta_i \ll 1$}
In this case the ion inertial length is much larger than the ion gyroradius, and we can neglect the dispersion corrections associated with the ion gyroradius (the finite Larmor radius effects). We thus assume $k_\perp \rho_i \ll 1$, but for the frequencies we now have $k_z v_{Ti}\ll \omega \ll k_z v_{Te}$. Other than that our consideration at this stage is general in that we do not require $\beta_e$, $kd_i$, or $k_z/k_\perp$ to be small. Our consideration here is different from previous studies of low-beta Alfv\'en dispersion relations \cite[e.g.,][]{lysak1996,hollweg1999}, where the hydrodynamic approximation $\omega\ll \Omega_i$ was made, and the ion inertial corrections proportional to $kd_i$ were therefore neglected. The dielectric tensor now has the form:
\begin{eqnarray}
\label{exx2} \epsilon_{xx}&\approx & -\frac{\omega^2_{pi}}{\omega^2-\Omega^2_i},\\
\label{eyy2}\epsilon_{yy}& \approx & -\frac{\omega^2_{pi}}{\omega^2-\Omega^2_i},\\ 
\label{exy2}\epsilon_{xy}&= &-\epsilon_{yx}\approx -i\frac{\omega\omega^2_{pi}}{\Omega_i(\omega^2-\Omega_i^2)},\\
\label{exz2}\epsilon_{xz}&= &\epsilon_{zx}\approx 0,\\ 
\label{eyz2}\epsilon_{yz}&= &-\epsilon_{zy} \approx  -i\frac{\omega_{pi}^2k_\perp}{\omega\Omega_ik_z},\\
\label{ezz2}\epsilon_{zz}&\approx & \frac{\omega_{pe}^2}{k_z^2v_{Te}^2}
-\frac{\omega^2_{pi}}{\omega^2}\left[1+\frac{3k_z^2 v_{Ti}^2}{\omega^2} \right],
\label{ezz}
\end{eqnarray}
where in $\epsilon_{zz}$ we kept the correction provided by the ion temperature, which will be needed for the dispersion of the ion-acoustic modes in section~4, but can be neglected for now.  As a result we obtain the general dispersion relation for a low-$\beta_i$ plasma:
\begin{eqnarray} 
\left(\omega^2-k_z^2v_s^2\right)\left[ \left(\omega^2-k^2v_A^2\right)\left(\omega^2-k_z^2v_A^2 \right)-\omega^2 k^2v_A^2k_z^2d_i^2 \right] \nonumber \\
 -\omega^2k_\perp^2 v_s^2\left(\omega^2-k_z^2v_A^2\right)+\omega^2k_\perp^2v_s^2k^2v_A^2k_z^2d_i^2=0.\quad
\label{dispersion_final}
\end{eqnarray}
As noted in \cite[][]{kuvshinov1994,ohsaki_m2004,hirose_etal2004}, equation (\ref{dispersion_final}) can also be derived from the Hall MHD with cold ions, $T_i\to 0$, however, in order to keep the finite ion-temperature effects, one needs to use the kinetic derivation. The terms containing $d_i$  describe the dispersion related to the ion inertial length. 

The following two limiting forms are often discussed in the literature.  The first is the hydrodynamic limit $\omega\ll\Omega_i$. In this case the terms containing $d_i$ can be neglected, and we recover the well known dispersion relations for the Alfv\'en, fast and slow modes. This limit is however not of importance for us since it neglects the dispersion.  

The second limit is the low electron beta case $\beta_e\ll 1$. In this limit we can neglect the terms containing the ion acoustic speed, that is, the second line in (\ref{dispersion_final}). As a result, the low-frequency ion acoustic mode $\omega^2=k_z^2v_s^2$ decouples,\footnote{The ion-acoustic mode is Landau damped unless  $T_e \gtrsim 6 T_i$, which limits $\beta_e$ from below.} and from the remaining equation we can derive the dispersive Alfv\'en and fast modes.

Here we consider the case of oblique propagation $k_z^2/k_\perp^2 \ll 1$. We do not make any assumptions about $\beta_e$ though. We start with the case when the propagation is oblique enough so that $k^2_z/k^2_\perp < \beta_e$. (If the opposite inequality holds we are back to the previous limiting case of small $\beta_e$, we will discuss this case later.)   In this limit the fast mode has high frequency, and it decouples from the other two modes. The equation for the fast mode is obtained if we approximate $\omega^2-k_z^2v_A^2\approx \omega^2$ and $\omega^2-k_z^2v_s^2\approx \omega^2$ in (\ref{dispersion_final}), which gives:
\begin{eqnarray}
\omega^4-\omega^2\left[k_\perp^2v_A^2(1+k_z^2d_i^2)+k_\perp^2v_s^2 \right]+k_z^2 k_\perp^4v_s^2v_A^2d_i^2=0.\,\,\,\,
\end{eqnarray}
For $k_zd_i\ll 1+(v_s/v_A)^2$ the solution is the fast or magnetosonic mode:
\begin{eqnarray}
\omega^2=k_\perp^2(v_A^2+v_s^2),
\label{fast_low_beta}
\end{eqnarray}
while in the opposite case $k_zd_i\gg 1+(v_s/v_A)^2$ it transforms into the whistler mode:
\begin{eqnarray}
\omega^2={k_z^2 k_\perp^2v_A^2d_i^2}.
\end{eqnarray}

\begin{figure}[t!]
\includegraphics[width=0.97\columnwidth]{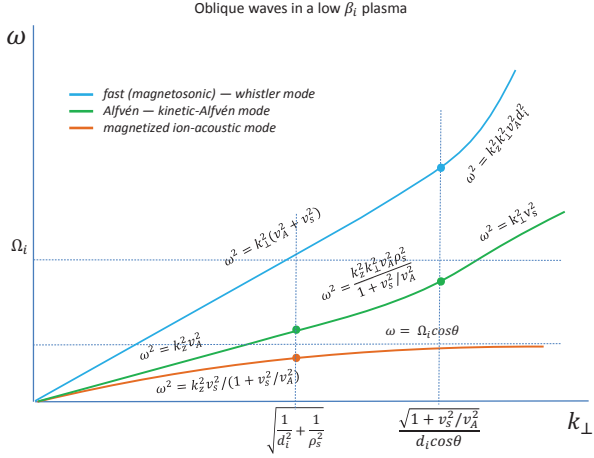}
\caption{\label{fig:curves} Sketch of dispersion relations for oblique waves in a collisionless low-$\beta_i$ plasma. The case of relatively high obliquity $k^2_z/k^2_\perp \ll \beta_e < 1$ is shown. The boundaries between different asymptotic regimes are marked by dots on the curves. }
\end{figure}

 The remaining dispersion equation for the two low-frequency modes is obtained if we approximate $\omega^2-k^2v_A^2\approx -k^2v_A^2$ in (\ref{dispersion_final}), which now turns into:
\begin{eqnarray}
&{~}& \omega^4\left[1+k_z^2d_i^2+v_s^2/v_A^2 \right]-\nonumber \\
& - &\omega^2\left[2k_z^2v_s^2+k_z^2v_A^2+k_z^2v_s^2k_\perp^2d_i^2 \right] +k_z^4v_s^2v_A^2=0.\qquad
\end{eqnarray}
In the long wave limit, when the $k_\perp^2 d_i^2$ containing term is negligible, the solutions are the shear-Alfv\'en mode and the ion-acoustic mode. At large $k_\perp^2 d_i^2$,  
the shear-Alfv\'en modes turns into the kinetic-Alfv\'en mode,
\begin{eqnarray}
\omega^2=\frac{k_z^2 k_\perp^2v_A^2\rho_s^2}{1+\frac{v_s^2}{v_A^2}},
\end{eqnarray}
while the ion-acoustic mode becomes:
\begin{eqnarray}
\omega^2=\Omega_i^2\cos^2\theta,
\end{eqnarray}
where $\theta$ is the angle between the wavevector and the background magnetic field. 
For each of the modes, the transition scale between the their long wave and short wave asymptotic solutions can be defined as the scale where the two solutions formally match. We thus obtain the transition scale for both the Alfv\'en and the ion-acoustic modes:
\begin{eqnarray} 
k_\perp^2 d_i^2 = 1+\frac{v_A^2}{v_s^2}.
\label{low_b_break}
\end{eqnarray}
A similar result can be obtained in the framework of Hall MHD, e.g., \cite[][]{schekochihin_etal2009}. The general behavior of these dispersion curves is schematically presented in Fig.~(\ref{fig:curves}). 

\begin{figure}[t]
\includegraphics[width=0.97\columnwidth]{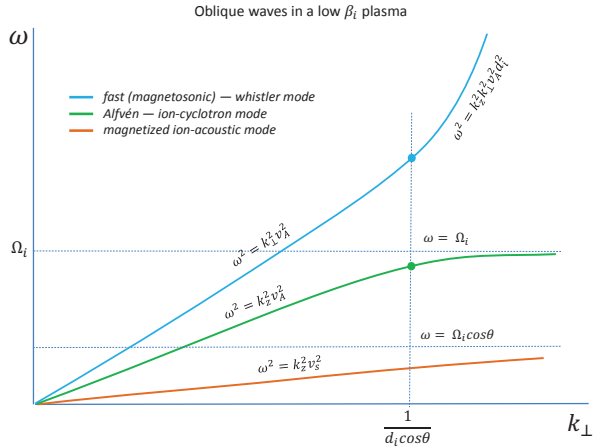}
\caption{\label{fig:curves_low_beta} Sketch of dispersion relations for oblique waves in a collisionless low-$\beta_i$ plasma. The case of relatively low obliquity $\beta_e \ll k^2_z/k^2_\perp \ll 1$ is shown.  The boundaries between different asymptotic regimes are marked by dots on the curves.}
\end{figure}

If the propagation is not very oblique so that $k_z^2/k_\perp^2>\beta_e$, then, as mentioned above, the dispersion relations can be found from the second limiting case. They have especially simple forms in the case  $k_z^2/k_\perp^2 \ll 1$, where we get for the fast and Alfv\'en modes, respectively
\begin{eqnarray}
\omega^2&=&k^2v_A^2(1+k_z^2d_i^2),
\label{very_low_b_break1}\\
\omega^2&=&k_z^2v_A^2/(1+k_z^2d_i^2).
\label{very_low_b_break2}
\end{eqnarray}
These dispersion relations are shown in~Fig.~(\ref{fig:curves_low_beta}). In both cases the breaks in the dispersion curves appear at $k_z\sim 1/d_i$. 

The obtained results suggest that for oblique propagation $k_z\ll k_\perp$, and small electron beta $1>\beta_e\gg \beta_i$, the spectral break should occur at the ion acoustic scale $\rho_s$, which is smaller than the ion inertial scale $d_i$. This contradicts the observations of \cite[][]{chen_etal2014}, where the break is observed at the ion inertial scale for such plasma parameters. In the following sections we analyze the physical processes that may explain this contradiction.

\section{Transition from Alfv\'en to kinetic-Alfv\'en turbulence in the strongly nonlinear regime}  
So far our consideration was based on the linear analysis. As a justification for that we note that in the case of oblique propagation, the turbulence in the solar wind is expected to be critically balanced. This means that the linear frequencies of interacting modes are comparable with the rates of nonlinear interaction at all scales.  Such assumption is based on phenomenological considerations  \cite[e.g.,][]{kraichnan65,goldreich_s95,boldyrev06,perez_etal2012}, and it is consistent with numerical studies of both Alfv\'en and kinetic-Alfv\'en turbulence \cite[e.g.,][]{cho_v00,tenbarge_h2012}. Qualitative properties of linear modes can therefore be present in such turbulence even in the strongly nonlinear regime \cite[][]{bale_etal2005,he_etal2012,howes_bk2012,klein_etal2012}. 
It may therefore be reasonable to relate the spectral breaks in the Alfv\'enic turbulence to the change in the dispersion properties of the associated linear modes.

In the limit $1\gg \beta_e \gg \beta_i$ the nonlinear dynamics of oblique kinetic-Alfv\'en waves is captured by the system of three equations for the electron density $n$, and the magnetic and velocity potentials, ${\bf b}_\perp={\hat z}\times \nabla \psi$, and  ${\bf v}_\perp={\hat z}\times \nabla \phi$ \cite[][]{camargo_etal1996,terry_mf2001}. Let us normalize the independent variables as follows,
$x' = x/L$, $t' = tv_A/L$, and the fields as 
${\phi}' = \phi/(Lv_A)$, ${n}' = \delta n/n_0$,  ${\psi}' = \psi/(B_0 L)$, where $L$ is the outer scale.  In the dimensionless variables the system takes the form (we omit the ${~}'$ signs):
\begin{eqnarray}
& {\partial_t} {\psi}=\nabla_\| {\phi} -{\frac{\beta_e}{2}}\frac{d_i}{L}\nabla_\|{n}, 
\label{psimod3}\\
& {\partial_t} { n}  + {\hat z}\times \nabla {\phi} \cdot \nabla {n}= \frac{d_i}{L}\nabla_\| \nabla_\perp^2\psi , 
\label{nmod3}\\
& \partial_t \nabla_\perp^2 \phi + {\hat z}\times \nabla \phi \cdot \nabla \nabla_\perp^2 \phi = \nabla_\| \nabla_\perp^2 \psi ,
\label{phimod3}
\end{eqnarray}
where we have denoted: $\nabla_\parallel = \partial_z + \boldsymbol{\hat{z}} \times \nabla_\perp {\psi}\cdot\nabla_\perp $.  
It is seen (by rescaling $n \to n/\sqrt{\beta_e/2}$, for example) that the only relevant microscale in the dynamics is $d_i\sqrt{\beta_e/2}=\rho_s$, the ion-acoustic scale. At larger scales, $k_\perp \rho_s \ll 1$, the equations for $\psi$ and $\phi$ turn into the reduced MHD equations that describe shear-Alfv\'en waves. The density decouples, and it is advected as a passive scalar. In the limit of small scales $k_\perp \rho_s \gg 1$, the velocity field decouples while density becomes dynamically important, and the resulting two-field system describes the kinetic-Alfven modes \cite[][]{boldyrev_p12}. 

To analyze the turbulence of Alfv\'{e}n waves we solve the above equations (\ref{psimod3}-\ref{phimod3}) in a periodic 3D rectangular domain using a pseudo-spectral code. The numerical procedure is analogous to that in \cite[][]{boldyrev_p12}.  The background magnetic field is along the $z$ axis. We supplement each of the equations with the random driving forces $f_\psi, f_n$, and $f_\phi$ that are applied at small $k$ in the Fourier space, satisfying $2\pi/L_\perp \leq k_{x,y} \leq 4\pi/L_\perp$ and $ 2\pi / L_z \leq k_z \leq 4\pi / L_z$. The Fourier components of the forces are drawn from a Gaussian distribution with zero mean and the variance chosen as to maintain same average amplitudes of the velocity, density and magnetic fields at the driving scale (the situation similar to observations). 
We choose ${v_A L_\perp}\sim {v_{\rm rms} L_z }$, so the turbulence is critically balanced at the outer scale. The average force refreshing time $\tau = 0.1 t_{A}$, where $t_{A}=L_z / v_A$. The number of grid points is the same in all directions, $N^3=1024^3$. 

\begin{figure}[t]
\includegraphics[width=\columnwidth]{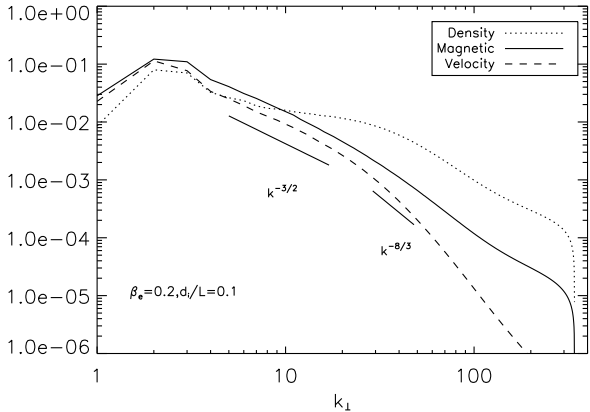}
\caption{\label{fig:no_vz_rev} The Fourier spectra of velocity, magnetic, and density fluctuations in Alfv\'enic turbulence driven at large scales, in a low-beta plasma. The inertial scale corresponds to $k_\perp=10$, the ion acoustic scale is $k_\perp\approx 30$. The transition from shear-Alfv\'en to kinetic-Alfv\'en turbulence is observed at the ion-acoustic scale.}
\end{figure}

The system keeps evolving until it reaches a steady state. We select the data from over $6 t_{A}$, where $t_{A}$ is Alfv\'{e}n crossing time. The interval of data sampling is $\sim 0.2 t_{A}$. The resulting spectra of strong oblique turbulence are shown in Fig.~\ref{fig:no_vz_rev}. 
It shows the spectral steepening indicating the transition from Alfv\'{e}n to kinetic-Alfv\'{e}n turbulence at the ion-acoustic scale $k_\perp \approx 30$. The density enhancement associated with increased compressibility of kinetic-Alfv\'{e}n modes, and the ``knee'' in the density spectrum are also observed at the same scale. The observed break in the spectrum of strongly nonlinear turbulence is less consistent with the ion-inertial scale,~$k_perp = 10$.

\section{Anomalous resistivity} 
In this section we discuss whether anomalous dissipation may be responsible for the spectral break in Alfv\'enic turbulence. In the case of low beta and highly nonisothermal plasma, $\beta_i \ll \beta_e < 1$, one may encounter a situation when the break in the spectrum is related to the nonlinear effect of anomalous resistivity. The physics of the phenomenon is the following~\cite[see, e.g.,][]{papadopoulos77}. As the scale of the Alfv\'en fluctuations decreases, the field-parallel electron current increases, since $\delta b_\lambda$ decreases slower than $\lambda$ in a turbulent spectrum. When the electron velocity associated with the current exceeds the phase velocity of the ion acoustic waves, these waves become unstable due to Cherenkov radiation. The interaction of the current with the waves then leads to the current damping, leading to the enhanced or anomalous resistivity. 

To estimate when this happens, one needs to retain the imaginary parts in the dielectric tensor for the ion acoustic waves. In a low-beta plasma, for the ion-acoustic waves we have $\omega \ll k_z v_A$ and $\omega\ll k_\perp v_A$. In this case, one can check that the $\epsilon_{xx}$ and $\epsilon_{zz}$ components of the dielectric tensor provide dominant contributions~\footnote{Another way to understand this fact is to note that the ion-acoustic mode is a potential mode, therefore, its dispersion relation can be derived from the Poisson equation $k_i\epsilon_{ij}k_j=0$, which transforms in the adopted reference frame, ${\bf k}=(k_\perp, 0,k_z)$, to the equation $k_\perp^2\epsilon_{xx}+k_z^2\epsilon_{zz}=0$.}.  The imaginary parts are small in $\epsilon_{xx}$; they should be retained only in the $\epsilon_{zz}$ component:   
\begin{eqnarray}
\delta\epsilon_{zz}=i\sqrt{\frac{\pi}{2}}\left[\frac{(\omega-k_zv_0)\omega_{pe}^2}{|k_z|^3v_{Te}^3}+\frac{\omega\omega^2_{pi}}{|k_z|^3v_{Ti}^3}\exp\left(-\frac{\omega^2}{2k_z^2v_{Ti}^2}\right)\right],
\label{anomalous}
\end{eqnarray}
where $v_0$ is the velocity of the electron current. For simplicity it is assumed to be constant.\footnote{Which may be a good approximation if the excited ion-acoustic modes have wavenumbers and frequencies exceeding those of the current.} The dispersion relation for the ion-acoustic wave with the ion-temperature corrections in (\ref{ezz}) has the form:
\begin{eqnarray}
\omega^2=\frac{k_z^2 v_s^2}{1+k_\perp^2 \rho_s^2}\left[1+3\frac{T_i}{T_e}(1+k_\perp^2\rho_s^2) \right],
\end{eqnarray}
where the second term in the square brackets is a small correction. 

It is easy to see from (\ref{anomalous}) that for sufficiently large $v_0$, the imaginary part of $\epsilon_{zz}$ becomes negative, leading to the instability. We consider separately the cases when the unstable modes have the wavelengths $k_\perp\rho_s\ll 1$ and $k_\perp\rho_s\gg 1$ \cite[cf.][]{papadopoulos77}. In the first case  the instability condition reads
\begin{eqnarray}
{v_0}>v_s\left[1+\left(\frac{m_iT_e^3}{m_eT_i^3} \right)^{1/2}\exp\left(-\frac{3}{2}-\frac{T_e}{2T_i} \right) \right].
\end{eqnarray}
This condition holds in the case of not very strong temperature difference, $T_e/T_i\lesssim 12$ for the hydrogen plasma (see below).  In the second case the instability condition is
\begin{eqnarray}
{v_0}>\frac{v_{Ti}}{k_\perp \rho_i}\left[1+\left(\frac{m_iT_e^3}{m_eT_i^3} \right)^{1/2}\exp\left(-\frac{3}{2}-\frac{1}{2k_\perp^2\rho_i^2} \right) \right].
\end{eqnarray}
The right hand side achieves minimum when 
\begin{eqnarray}
\frac{1}{(k_\perp \rho_i)^2}\approx { \ln\left[ \left(\frac{T_e}{T_i} \right)^{3}\frac{m_i}{m_e} \right]-{3}},
\end{eqnarray}
which together with $k_\perp \rho_s > 1$ implies for the hydrogen plasma $T_e/T_i\gtrsim 12$. The instability condition is then 
\begin{eqnarray}
v_0>v_{Ti}\sqrt{ \ln\left[ \left(\frac{T_e}{T_i} \right)^{3}\left(\frac{m_i}{m_e} \right)\right]-{3}},
\label{threshold}
\end{eqnarray}
which for $T_e/T_i\gtrsim 12$ gives $v_0\gtrsim 3.5 v_{Ti}$.

\begin{figure}[t]
\includegraphics[width=\columnwidth]{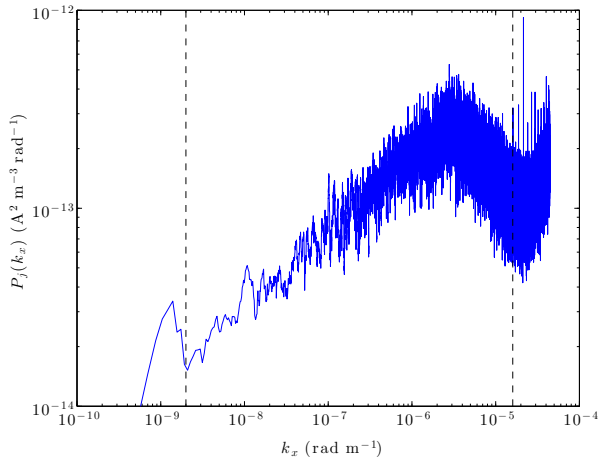}
\caption{\label{fig:current} Power spectral density of current density $P_j(k_x)$ for the solar wind interval in Figure 2a of \citet{chen_etal2014} calculated from the trace magnetic spectrum $P_B(k_x)$ as $P_j(k_x)=((\sqrt{2}k_x/\sin\theta_\mathrm{Bv})/\mu_0)^2 P_B(k_x)$, where $\sin\theta_\mathrm{Bv}$ is the angle between the mean magnetic field and the solar wind (sampling) direction, $k_x=2\pi f_\mathrm{sc}/v_\mathrm{sw}$ is the wavenumber in the sampling direction assuming the Taylor hypothesis, $f_\mathrm{sc}$ is the spacecraft-frame frequency and $v_\mathrm{sw}$ is the solar wind speed. The spectrum integrated between the black dashed lines (to exclude windowing effects at small $k_x$ and instrumental noise at large $k_x$) gives $\delta j^2=2.7\times 10^{-18}$ A$^2$ m$^{-4}$, which corresponds to a velocity $v_0=9.7$ km s$^{-1}$, a factor of 3.8 smaller than $v_s$.  }
\end{figure}

The electron velocity $v_0$ can be estimated from the field-parallel electron current $J_\|=n_0ev_0$.  The current is related to the magnetic field fluctuations: $i{\bf k}\times {\bf b}_k=(4\pi/c){\bf J}_k$. The rms current can therefore be found from the measured spectrum of magnetic fluctuations $J^2=(c/4\pi)^2\int k^2 E(k)dk$. Such a current can be estimated from the observational sets of \cite[][]{chen_etal2014}, see Fig.~\ref{fig:current}.
The estimates lead to the electron velocity $v_0$ about $4$ to $15$ times smaller then the ion-acoustic velocity for the measured plasma parameters. The current is however known to be highly intermittent in both MHD and kinetic turbulence \cite[][]{zhdankin_etal2013,wan_etal2012}, meaning that a significant fraction of the energy is dissipated in the regions where the current is several times larger than its rms value. At present it is unclear whether the current intermittency in the solar wind  is strong enough to account for the inferred mismatch between the rms electron drift velocity and the ion-acoustic speed. It is therefore hard to conclude  whether the anomalous resistivity can be essential at subproton scales.

\section{Interactions of Alfv\'en waves with ion-acoustic and fast fluctuations}
In a low-beta, $\beta_i\ll 1$, and non-isothermal, $T_i\ll T_e$, plasma a significant role in a turbulent cascade can be played by the ion-acoustic modes and the fast modes, where the density fluctuations are not small. Such modes are subject to strong Landau damping in the case $\beta_e\sim \beta_i \sim 1$, which partly explains rather low level of density fluctuations observed in the corresponding turbulence. In the non-isothermal, low-beta case, however, the density fluctuations associated with these compressible modes become dynamically essential. One of the important implications is the possibility of parallel energy cascade due to decay interactions of the Alfv\'en waves with the ion-acoustic and fast waves.

\subsection{Scatter of Alfv\'en waves by ion-acoustic fluctuations}
For an estimate we consider weak turbulence of shear-Alfv\'en and ion-acoustic modes \cite[excellent accounts of this theory can be found in, e.g.,][]{kuznetsov2001,chandran2008}. In the absence of the  ion-acoustic modes, the turbulence of shear-Alfv\'en modes is known to transfer the energy in the field-perpendicular direction, populating mostly wavemodes with $k_\perp \gg k_\|$. The rate of nonlinear energy transfer due to Alfv\'en wave interactions is estimated as $\gamma_A \sim k_\perp^2 v^2/k_\| v_A $ \cite[e.g.,][]{ng_b96}, where $v$ is the amplitude of the velocity fluctuations in the wave. This expression is applicable also in the case of strong turbulence, if one assumes that the critical balance condition $v/v_A\sim k_\|/k_\perp$ is satisfied \cite[e.g.,][]{goldreich_s95}.  

\begin{figure}[t]
\includegraphics[width=\columnwidth]{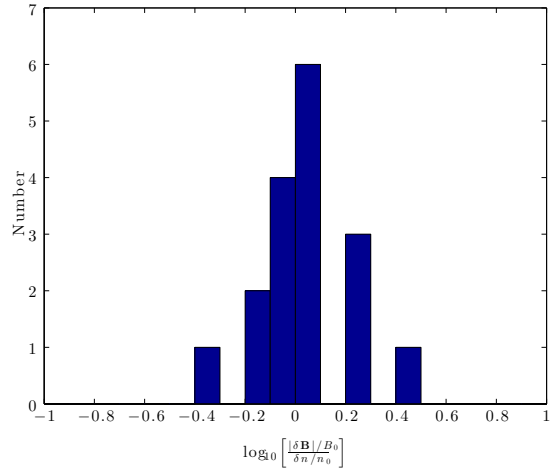}
\caption{\label{fig:histogram2} Histogram of the ratio of normalized magnetic field fluctuations to ion density fluctuations for the low beta solar wind intervals discussed in \citet{chen_etal2014} in the range of spacecraft-frame frequencies $3\times 10^{-3}$ Hz to $1\times 10^{-2}$ Hz (within the MHD inertial range). Absolute values of density fluctuations are used.}
\end{figure}

The Alfv\'en wave interaction with the ion-acoustic fluctuations is dominated by scattering of the Alfv\'en waves by the density inhomogeneities, as the magnetic fluctuations are small in the low-beta ion acoustic modes. 
In the Elsasser variables, ${\bf z}^{\pm}={\bf v}\pm {\bf b}/\sqrt{4\pi \rho}$, the MHD equations take the form \cite[e.g.,][]{marsch1987}:
\begin{eqnarray}
& \partial_t z^{\pm} \mp (v_A\cdot \nabla) z^{\pm}+ (z^{\mp}\cdot \nabla)z^{\pm} =-\frac{1}{\rho}\nabla p -\nonumber \\
&-v_A\nabla\cdot v+ \frac{1}{2}(z^--z^+) \nabla \cdot (b/\sqrt{4\pi\rho}+v/2) +\nonumber \\
& +\frac{1}{2}(z^--z^+) \nabla \cdot v_A, \label{elsasser} \\
&\partial_t \rho +\nabla \cdot \rho v =0, 
\end{eqnarray} 
where $v_A=B_0/\sqrt{4\pi\rho}$, and the pressure $p$ includes both the kinetic and magnetic parts.   The interaction of the shear-Alfv\'en waves with the density fluctuations is described to the leading order by the last term on the right hand side of (\ref{elsasser}). 
If the majority of density fluctuations have $k_\|\gg k_\perp$, the energy of  Alfv\'en waves is transferred in the $k$ space mostly along the field-parallel direction. Let us assume that an Alfv\'en wave, say $z^+$, propagates along the background magnetic field through a series of density humps of typical scale $l$, having intensities $\delta n_l$. 

Following a standard weak-turbulence consideration, we estimate that in each crossing time $\sim l/v_A$ the amplitude of the wave is distorted by $\delta z^+\sim (\delta n_l/n)z^+$. Since these ampitude changes are random, one needs to accumulate $\sim (n/\delta n_l)^2$ distortions to change the amplitude significantly. During an interaction the frequency of the wave changes by $\Delta \omega\approx \pm \omega \sqrt{2\beta_e}$, with the sign depending on the mutual directions of propagation of the Alfv\'en and ion-acoustic perturbations. The energy diffusion in the phase space is thus impeded by a factor $(\Delta \omega/\omega)^2\sim \beta_e$. The inverse time of energy cascade in the phase space can thus be astimated as $\gamma_S= \beta_e (v_A/l)(\delta n_l/n)^2$. 

Parenthetically, we note that this consideration may need to be modified if the density fluctuations are allowed to steepen and develop shocks. In this case, we may assume that the $z^+$ wave is propagating through a set of density discontinuities of amplitude $\delta n_l$ separated by distance $l$. In each crossing time $\sim l/v_A$, the wave looses its energy; a fraction of energy transferred to the reflected $z^-$ wave is equal to $\delta (z^+)^2/(z^+)^2\sim (\delta n_l/n)^2$.  The frequencies of the reflected waves differ by $\Delta \omega\approx \pm \omega \sqrt{2\beta_e}$ from the frequency of the propagating wave. 
Since the signs of individual frequency shifts are random, one needs to accumulate $(\omega/\Delta\omega)^2\sim 1/(2\beta_e)$ subsequent reflections  before the frequency changes by its magnitude. Therefore, the time of the frequency shift is $\delta \tau\sim (l/v_A)/(2\beta_e)$. During this time, a fraction of energy $(\delta n_l/n)^2$ will cascade to smaller scales. The rate of energy transfer is therefore $\gamma_S=[\delta (z^+)^2/\delta \tau]/(z^+)^2\sim \beta_e (v_A/l)(\delta n_l/n)^2$, which provides the same estimate as before.   

Comparing the rates of the energy transfer, we get:
\begin{eqnarray}
\gamma_S/\gamma_A\sim \beta_e (k_\|/k_\perp)^2(\delta n/n)^2(\delta b/B_0)^{-2},
\end{eqnarray}
where we have substituted $l\sim 1/k_\|$. Observations \cite[][]{chen_etal2014} demonstrate that in the regime $1\gg \beta_e \gg \beta_i$, the density and magnetic fluctuations are on the same order  in the MHD inertial interval, $\delta n/n \sim \delta b/B_0$, see Fig.~\ref{fig:histogram2}, therefore the Alfv\'enic and density fluctuations exhibit mostly field-parallel energy cascade inside the cone $k_\|/k_\perp \gtrsim 1/\sqrt{\beta_e}$. Assuming that the velocity and magnetic fluctuations have the same turbulence spectra, we then obtain from the constant rate of energy cascade, $\epsilon\sim \gamma z^2\sim {\rm const}$, the energy spectrum
\begin{eqnarray}
E(k_\|)dk_\|\propto k_\|^{-3/2}dk_\|. 
\end{eqnarray}
This result agrees with the analytic consideration of \cite[][]{kuznetsov2001}, if we assume that the propagation is mostly parallel. According to the dispersion relation (\ref{very_low_b_break2}) in this case the cascade will have a break point at $k\sim k_\|=1/d_i$. This may be consistent with the observations in \cite[][]{chen_etal2014}.

\subsection{Interactions of Alfv\'en waves with fast modes} 
An efficient energy transfer in the parallel direction can also be provided by the interaction with the fast modes. These modes have comparable normalized magnetic and density fluctuations, therefore, the interaction is described by the last term in the left-hand side and by the last two terms in the right-hand side of~(\ref{elsasser}). The energy transfer is isotropic in the k-space, with the rate estimated as $\gamma_F\sim (v_A/\lambda)(\delta n/n)^2$, where $\lambda\sim 1/k$. This rate is comparable to the rate of field-perpendicular energy transfer due to the shear-Alfv\'en modes. Therefore, interactions with fast modes provide an efficient energy transfer in the field-parallel direction, and can explain the spectral break at $k\sim k_\|\sim 1/d_i$. The energy spectrum, estimated from the constant energy flux $\gamma_Fz^2={\rm const}$, is given by
\begin{eqnarray}
E(k)dk\propto k^{-3/2}dk,
\end{eqnarray}
which is also consistent with the analysis of \cite[][]{chandran2005,cho2002}, and may agree with the  observations of \citet{chen_etal2014}.  
This consideration holds except for the propagation directions close to the background magnetic field, where the density and the field-parallel magnetic fluctuations of the fast mode become small, $\delta n/n=\delta b_z/B_0=(k_\perp/k)\delta b/B_0$. At the propagation angles $k_\perp/k\lesssim \sqrt{\beta_e}$, the energy transfer is then dominated by the scattering of the shear-Alfv\'en waves by the ion-acoustic modes, as described in the previous section.

\section{Observational evidence for dynamically important density fluctuations} 
\begin{figure}[t]
\includegraphics[width=\columnwidth]{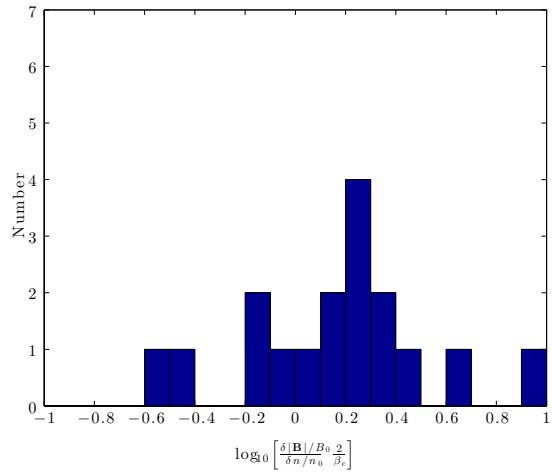}
\caption{\label{fig:histogram1} Histogram of the ratio of normalized {\em magnetic field strength} fluctuations to ion density fluctuations for the low beta solar wind intervals discussed in \citet{chen_etal2014} in the range of spacecraft-frame frequencies $3\times 10^{-3}$ Hz to $1\times 10^{-2}$ Hz (within the MHD inertial range). Note deviations from the pressure balance condition. Lower values of the argument are consistent with the presence of ion-acoustic modes, higher values -- with the presence of fast modes. Absolute values of fluctuations are used.}
\end{figure}

The measurements of spectral anisotropy are not available for the low-beta intervals analyzed in  \cite[][]{chen_etal2014}. We can however infer the significance of density fluctuations in the energy cascade in a low-beta plasma from comparison of the density fluctuations with the fluctuations of magnetic field strength. 
If the density fluctuations are not dynamically important, e.g., they are passively advected by the  shear-Alfv\'en turbulence \cite[e.g.,][]{chandran2008,howes_bk2012}, they satisfy locally the pressure-balance condition $\delta |B|/B_0=-(\beta_e/2)\delta n/n_0$. In this case, the fluctuations of the magnetic field are dominated by the field-perpendicular component, $\delta b_\perp \gg \delta b_z$, while the small field-parallel fluctuations are related to the fluctuations of the magnetic field strength $\delta |B| \approx \delta b_z$.  The observations however show that the pressure-balance condition is not generally observed in the considered case, see Fig.~\ref{fig:histogram1}. The deviations from the pressure balance may be explained by the presence of the fast modes, for which $\delta |B|/B_0=\delta n/n_0$, and/or almost field-parallel ion-acoustic modes, for which $\delta |B|/B_0=-(k_\perp/k)^2(\beta_e/2)\delta n/n_0$, which lends support to the importance of these modes in the turbulent cascade, as discussed in the previous sections.  
{~}

\section{Conclusions}  
We propose that according to the critical balance condition, the breaks in the turbulent energy spectra of a weakly collisional plasma, as those observed in the solar wind and in the interstellar plasma \cite[e.g.,][]{bale_etal2005,bourouaine2012,markovskii2008,haverkorn_s2013}, are related to the breaks in the corresponding dispersion relations. For a high beta plasma, $\beta_i\gg 1$, the  break for the shear-Alfv\'en mode occurs at scales $k_\perp \sim 1/\rho_i$.
For a low beta plasma $\beta_i\ll 1$, on the other hand, the situation crucially depends on the obliquity of propagation, and the electron plasma beta $\beta_e$.  

If $k_z^2/k_\perp^2<\beta_e$, the turbulent energy cascade is predominantly in the field-perpendicular direction. The energy transfer is dominated by interactions among the shear-Alfv\'en waves, and the spectral break for the shear-Alfv\'en mode depends on the combination of the ion-acoustic scale and the ion inertial length scale, as follows from formula (\ref{low_b_break}). For $\beta_e < 1$ and $k_z\gtrsim k_\perp$ the the energy cascade governed by interactions of shear-Alfv\'en modes with fast modes can compete with the cascade governed by mutual shear-Alfv\'en interactions. In this case significant fractions of energy can cascade in the direction of large $k_\|$. Finally, for $k_z^2/k_\perp^2 > 1/{\beta_e}$ the energy cascade may be predominantly in the direction of large $k_\|$, and it may be dominated by interactions of shear-Alfv\'en waves with the ion-acoustic modes. In the last two cases, a spectral break would appear at the scale $k \sim 1/d_i$, in agreement with the recent observations \cite[][]{chen_etal2014}.

Our analysis, together with the observations of \cite[][]{chen_etal2014}, may provide an evidence for the significance of a field-parallel turbulent cascade in a low beta plasma. They also emphasize the importance of distinguishing among various plasma regimes in describing turbulence in a collisionless plasma, and in interpreting observational results.


\begin{acknowledgments}
This work was supported by the NASA grant NNX14AH16G, US DoE grant DE-SC0003888, and the NSF Center for Magnetic Self-organization in Laboratory and Astrophysical Plasmas at U. Wisconsin-Madison. This research was supported in part by the National Science Foundation under Grants No. NSF PHY11-25915 and NSF AGS-1261659; SB appreciates the hospitality and support of the Kavli Institute for Theoretical Physics, University of California, Santa Barbara, where part of this work was conducted. C. H. K. Chen is supported by an Imperial College Junior Research Fellowship. The numerical simulations were made possible through the NSF TeraGrid allocation TG-PHY120042.
\end{acknowledgments}

\bibliographystyle{apj}

\end{document}